**Structural tuning of reduced exciton mass in layered HOIP compounds: Causation vs. correlation**

*Isaac R. Burkholder, Cindy Y. Wong, André Schleife, Kameron R. Hansen, John S. Colton, Branton J. Campbell**

I. R. Burkholder, J. S. Colton, B. J. Campbell
Department of Physics and Astronomy, Brigham Young University, Provo, UT 84602, USA
E-mail: branton@byu.edu

C. Y. Wong, A. Schleife
Department of Materials Science and Engineering and Materials Research Laboratory, University of Illinois at Urbana-Champaign, Urbana, IL 61801, USA

K. R. Hansen
Department of Chemical and Biochemical Engineering, University of Iowa, Iowa City, Iowa 55242, USA

Keywords: hybrid organic-inorganic perovskite, structure-property relation, exciton engineering, symmetry mode analysis, irreducible representation

**Reduced exciton mass ($\mu$) was recently reported to correlate strongly with a framework distortion in a series of nine single-layer (2D) metal-halide perovskite (HOIP) compounds. Specifically, $\mu$ was observed to increase in tandem with an alternating $PbI_4$ octahedral tilt about an in-plane axis. In this work, we use group representation theory to decompose the observed framework distortions into displacive symmetry modes of a common high-symmetry parent framework. We find that all nine distorted frameworks involve linear combinations of the same six symmetry modes, which have been reported to contribute to the framework distortions of a wide range of HOIP compounds. We show that these modes have highly correlated impacts on the band structure. To differentiate causation from correlation, we vary the amplitude of each mode independently and use density-functional theory to determine the resulting electronic band structures, from which $\mu$ is extracted. We find that bond-transverse displacements of the equatorial halide atoms increase $\mu$, while bond-transverse displacements of the apical halide atoms decrease it. Bond-axis displacements appear to have little or no effect on $\mu$. Our results demonstrate three new structure-property relationships, revealing a promising new avenue for exciton engineering in layered perovskite materials.**

## 1. Introduction

Single-layer ($n$=1) hybrid organic-inorganic perovskite (HOIP) semiconductors have inorganic perovskite layers consisting of corner-sharing metal-halide ($BX_6$) octahedra separated by layers of monovalent or divalent organic molecules (B), for an overall chemical formula of the form $ABX_4$ or $A_2BX_4$. In-plane halides that are shared between octahedra are referred to here as "equatorial halides", and out-of-plane halides which are part of single octahedron are referred to as "apical halides." HOIPs have made impressive progress in technological applications in recent years. Steady improvements in solar cell materials and processes, for example, have roughly doubled power conversion efficiency over the past decade.[1] Similar progress has been made in applications involving LEDs, lasers, and spintronics.[2–4]

A central feature contributing to the importance of HOIP materials is the extensive tunability of their optical and electronic properties, which is made possible by a seemingly infinite variety of organic spacer molecules. The electronic band structure of a HOIP material is primarily dominated by the inorganic framework,[5–8] which becomes distorted through its interactions with the organic spacers. Now that many hundreds of different organic molecules can be combined with dozens of different framework chemistries, the engineering of framework distortions that invoke a controlled effect on the band structure is emerging as a pressing problem.

Robinson *et al.* in 1971 characterized octahedral and tetrahedral distortions in $(Mg,Fe)_2SiO_4$ in terms of the standard deviation of the bond-angle distribution and a quadratic elongation parameter related to the standard deviation of the bond-length distribution.[9] This approach, which effectively reduces the complexity of a distorted polyhedron from three parameters per vertex down to a total of two parameters, has been applied to several HOIP framework distortions.[10,11] Complexity reduction can be useful in detecting the presence of a structure-property relationship when the detailed mechanism of the relationship can't be readily discerned. However, distinguishing causation from correlation is only possible when each structural variable is considered independently.

A wealth of previous work has demonstrated correlations between long-range HOIP framework distortions and important electronic band-structure features, such as band gap ($E_g$), exciton binding energy, reduced exciton mass ($\mu$), ionization energy, electron affinity, Rashba-Dresselhaus spin-splitting, chirality-induced spin-selectivity, etc.[4,5,10,12–21] One of the first observed trends was for a series of HOIP compounds based on Sn-I frameworks and fluorinated phenethylamine cations, where Knutson *et al.*[5] reported a correlation between $E_g$ and a metal-halide-metal (X-B-X) angle characteristic of octahedral tilting. Similar relationships have been observed in other compounds.[10-13,15–19,22-26]

The X-B-X bond angle defined by Knutson *et al.*[5] measures the projection of a specific X-B-X bond angle in three dimensions onto the two-dimensional plane of the framework sheets. Other complexity-reducing quantifiers of HOIP framework distortions have proliferated in the literature in recent years,[4,13–17,24,26,27] each involving some combination of bond angles and/or bond lengths; many are loosely related to the X-B-X angle of Knutson *et al.*[5] For example, Hansen *et al.* defined an octahedral tilting angle, referred to as $|\beta - \beta'|$, which roughly quantifies the amount of corrugation in the inorganic framework layers.[16] While quantifiers like those of Knutsen *et al.* and Hansen *et al.* were important advances, we see that they have two primary weaknesses. First, they neglect other correlated structural variables which also have the potential to influence $E_g$. Second, most HOIP frameworks have multiple symmetry-distinct metal sites, each of which has multiple distinct bond angles, any of which can be projected onto the plane, so that the choice of angle is somewhat arbitrary. While the observed changes to the band-structure can often be explained in terms of hybridization between the metal and halide orbitals,[5-7,15,17,22,23,25,26] such approaches still fundamentally rely on proper structural descriptors to connect the orbital-hybridization explanation to the long-range structure-property trends.

Local structure features have recently been shown to strongly influence the electronic and phononic band structures of HOIP materials.[28-30] In particular, Zhao *et al*. showed that several well studied 3D perovskites, such as $CsPbI_3$, $CsSnI_3$, $MAPbI_3$, $FAPbI_3$, and $FASnI_3$, have lower free energies when allowed to relax within a large supercell such that multiple distinct octahedral geometries arise across the supercell, leading them to describe their models as polymorphic.[28] Though such long-range superstructures have not been observed experimentally, analogous local-structure models were demonstrated to improve fits to neutron pair distribution function data. Their polymorphic models also yielded DFT-calculated bandgaps, and dielectric constants, and formation enthalpies that better match experiments. Zacharias *et al*.[29] adapted polymorphism 3D-perovskite models to a high-throughput scheme and showed that the band gap and effective mass depend on local B-X bond length and local B-X-B bond-angle, similar to long-range dependencies that have been reported.[5,10,14-17,19] Local structure effects on band structure have also been reported in a 2D HOIP perovskite: in single-layered $(3AMP)PbI_4$, a local displacive modulation was correlated with an increase in the reduced exciton mass.[30]

With the growing interest in structure-property relationships in HOIP compounds, there is a pressing need for a robust and self-consistent parameter set for the space of structural distortions common to an entire structure type. Such a common parameter set arises naturally from group representation theory, and can simultaneously accommodate displacive, occupational, magnetic, rotational, and lattice-strain variables. A group-theoretic parameter set spans the space of all possible distortions of all single-layer (2D) HOIP frameworks. We refer to these parameters as symmetry modes. Symmetry-mode analysis (SMA) characterizes a low-symmetry (distorted) child structure relative to a high-symmetry (undistorted) parent structure in terms of how the child differs from the parent.[31,32] Each symmetry mode belongs to an irreducible matrix representation (irrep) of the parent space group and communicates a specific way in which a child can differ from its parent. All of the complete space-group irreps have been tabulated.[33] Unlike bond-angle methods, there is no built-in complexity reduction; the symmetry-mode basis is completely general. By decomposing a distorted child HOIP framework into a set of symmetry mode amplitudes, distortions can be compared across structures in a robust and unambiguous fashion. It is common and often useful to apply symmetry analysis in a lesser sense by relating simple ad hoc structural descriptors to a physical-property tensor component, band-structure feature, or spectral feature.[3-6,10-19,22,23,26,34,35] When doing so, one should remember that simple descriptors tend to be linear superpositions of correlated but distinct symmetry modes. [31,32,36-42]

SMA has previously been used to characterize framework distortions in layered perovskite frameworks of the oxide and halide varieties.[36-43] Examples involving 3D oxide perovskites are too numerous to list here. McNulty and Lightfoot explored the irreps corresponding to symmetry-allowed cooperative framework-tilt patterns in over 250 single-layer lead-halide perovskites and tabulated those responsible for significant distortions.[44] Liu *et al.* identified 47 octahedral tilt-mode patterns in single-layer HOIP frameworks derived from an idealized parent framework of the Ruddlesden-Popper (RP) layer-stacking type.[45] Primary distortions within a 2×2×2 in-plane supercell of an RP framework lead to 47 distinct isotropy subgroups, each of which involves a joint order parameter based on some superposition of irreps capable of cooperative tilt patterns (also called rigid unit modes). They then examined 140 single-layer HOIP compounds based on a variety of different framework compositions and reported on the frequency with which each type of tilt-mode pattern is observed.[45] The symmetry-mode parameter set is also ideal for local-structure models,[46,47] and has been employed successfully for a 3D HOIP perovskite material.[48] In any case, our present focus is on the dependence of band-structure on long-range structural distortions.

SMA provides an especially powerful distortion basis when combined with density functional theory (DFT) calculations, which allow one to unambiguously relate direct-space distortions to features of the electronic band structure.[3,40,49,50] In the present work, we employ symmetry-mode parameters and DFT calculations to isolate the parameter(s) responsible for the reported relationship between reduced exciton mass and octahedral-tilting in a collection of 2D HOIPs,[16] and thereby differentiate causation from correlation. We vary each symmetry mode in isolation around its experimentally observed value and use DFT to obtain $\mu$. The exciton's reduced effective mass has a direct effect on important optoelectronic HOIP properties like carrier mobility, exciton binding energy, diffusion length, and even the band gap energy,[15-16] which are of practical importance in designing materials for LED, solar cell, laser, and other semiconductor applications.

## 2. Experimental Section/Methods

To perform displacive symmetry-mode analysis (SMA) on the inorganic framework of each child structure, we use the 'mode decomposition' feature of the web-based ISODISTORT[32] program of the ISOTROPY Software Suite (iso.byu.edu) after pre-processing the relevant CIF structure file with ISOCIF. Automatic relative basis and origin detection was successful in all but two cases (EAOH and C14), for which we manually determined the basis and origin of the child relative to the parent (in unitless parent coordinates). We performed SMA runs in batch mode using a python script that communicated directly with the web server and saved the 'all modes details' output from ISODISTORT for each structure. Descriptions and depictions of the atomic displacements involved in each of the found symmetry modes are given in Section 3.4 below.

All DFT calculations in this study were performed using the Vienna Ab-initio Simulation Package (VASP) code with the projector augmented wave method.[51,52] The exchange-correlation functional is described by the PBEsol generalized gradient approximation (GGA).[53] Calculations in this work were done using a Γ-centered 6×6×6 *k*-point grid, while calculations taken from Hansen *et al.*[16] used the k-point grids shown in **Table 1**, with a plane-wave energy cutoff of 600 eV. Effective electron and hole mass tensors were calculated using the effective mass calculator (EMC) at the conduction band minima and valence band maxima of the DFT-PBEsol band structure.[54] To calculate the reduced exciton mass $\mu$, we use the following equation:

$$1/\mu = 1/m_e + 1/m_h$$

where $m_e$ and $m_h$ are the in-plane electron and hole effective mass, respectively. Hansen *et al.*[16] used a more sophisticated hybrid functional (HSE06) for their DFT calculations, which achieved greater overall accuracy. While there is the expected underestimation of the band gap from using PBE compared to HSE, the trends we observe between the structure and effective mass are expected to be consistent for the two methods. Therefore, we use PBE instead of HSE to reduce the otherwise exceedingly high computational cost.

**Table 1.** K-*point grids used for each structure in Hansen* et al.[16] *The chemical formula and CCDC identifier for each structure are found in Table 2.*

| Structure name | K-point grid |
|---|---|
| $(BA)_2$@293K | 3x3x1 |
| $(BA)_2$@100K | 3x3x1 |
| C7 | 2x2x1 |
| C10@172K | 3x2x1 |
| C10@268K | 5x5x1 |
| C14 | 7x7x1 |
| 1F-PEPI | 2x2x4 |
| $(PEA)_2$ | 2x2x1 |
| EAOH | 4x4x4 |

## 3. Results and Discussion

Untangling the impacts of multiple correlated structural distortions on the reduced exciton mass requires a robust-yet-simple parameter set. In this section, we first describe a high-symmetry parent structure type that provides a common parameter set for the symmetry-mode analysis of one framework layer from any single-layer HOIP. We then explain the selection of the nine experimental HOIP structures analyzed here. Following a brief discussion of the preparation of child framework structures and their corresponding topological parent structures, we describe the SMA results in detail, including the geometric nature of each *structurally-active* symmetry mode (i.e. one with a significantly non-zero amplitude) and its contribution to each child structure. Finally, we employ DFT to probe the effect of each structurally-active symmetry mode on reduced exciton mass and conclusively identify the mode patterns with the greatest impact.

### 3.1. Parent structures for SMA

Whilst any higher-symmetry structure with the same bonding topology and chemical composition can serve as a parent structure for SMA, it is generally most informative to use the highest-symmetry parent structure achievable by enforcing approximate symmetries (a.k.a. pseudosymmetries) without modifying the atomic bonding topology. We often call this the *topological parent structure* and refer to its space group as the topological parent space group (TPSG).[3]

For a structure where multiple distinct layer stackings are possible, the topological parent structure may not be unique. For layered HOIP frameworks, there are three inequivalent high-symmetry stackings of the idealized perovskite layer, as seen in **Figure 1**, each of which might be chosen as the topological parent:[44,45]

1) The Dion-Jacobson (DJ) stacking has in-plane interlayer offset (0, 0), resulting in TPSG *P*4/*mmm* (#123), Figure 1(left).
2) The Ruddlesden-Popper (RP) stacking has in-plane interlayer offset (1/2, 1/2), resulting in TPSG *I*4/*mmm* (#139), Figure 1(middle).
3) The Half-Zero (H0) stacking has in-plane interlayer offset (1/2, 0), resulting in TPSG *Cmmm* (#65), Figure 1(right). McNulty and Lightfoot previously used the terminology "DJ2" for this structure type,[44] whereas Liu *et al*. used the word "intermediate" to express the fact that the H0 offset lies between those of DJ and RP.[45]

For layered HOIP materials, it has recently become common to interpret the DJ and RP labels in terms of molecular stoichiometry ($A_1PbI_4$ for DJ and $A_2PbI_4$ for RP) rather than stacking type, though we do not employ such usage here. The recent evolution in terminology is likely due to the fact that the two interpretations very often (but not always) align.

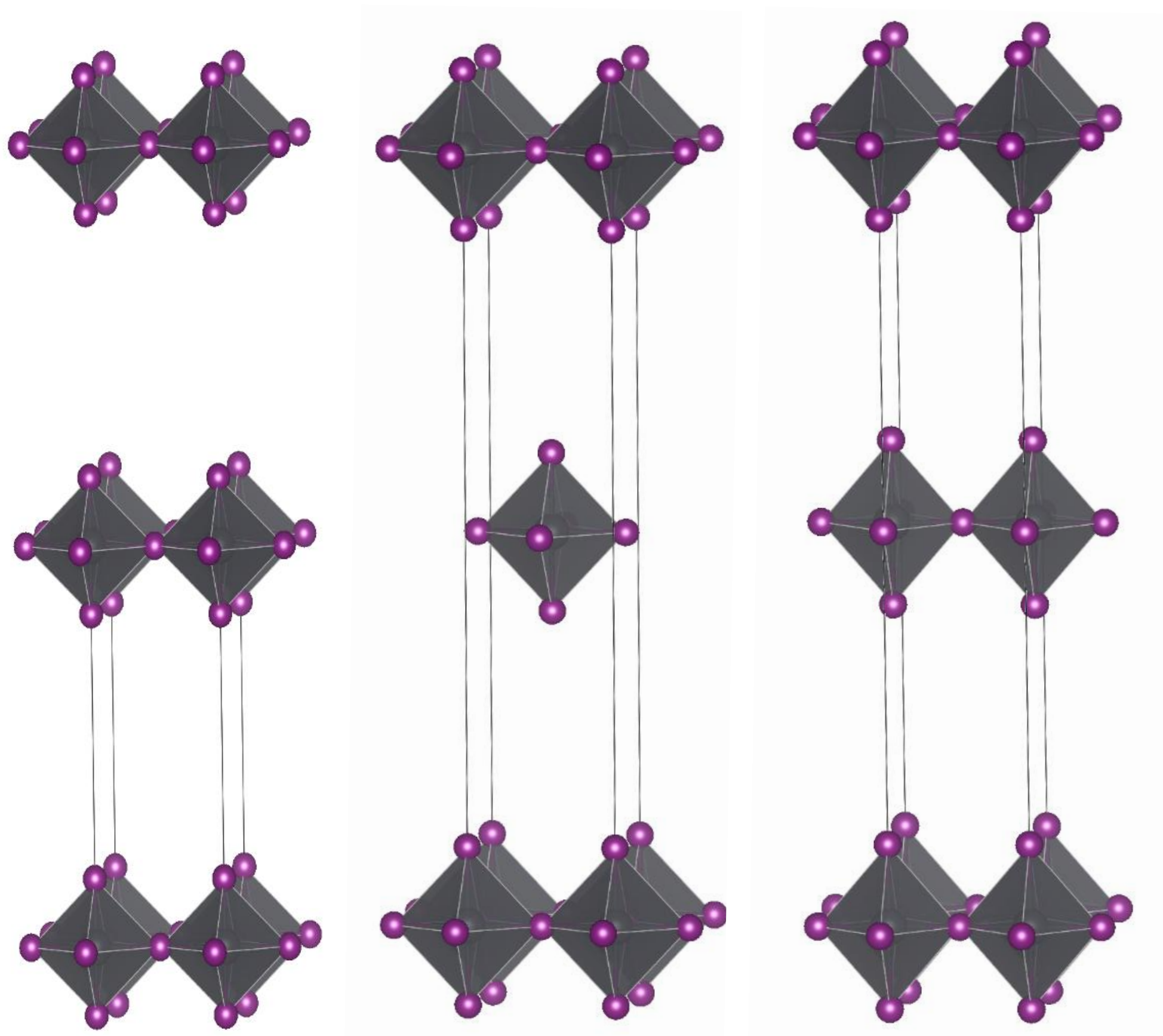

**Figure 1.** *Examples of topological parent structures for n=1 2D HOIPs. From left to right: Dion-Jacobson (DJ, P*4/*mmm, #123), Ruddlesden-Popper (RP, I*4/*mmm, #139), and Half-Zero (H0, Cmmm, #65), with boxes indicating the conventional unit cell. All structures in this study were analyzed using the DJ topological parent structure.*

In the present work, we specifically explore distortions that occur within a single HOIP framework layer and ignore any interlayer correlations. Within this analytical scope, the actual layer stacking is not relevant. Thus, regardless of the actual stacking pattern of a distorted child framework, we artificially shift the interlayer-stacking offset into a DJ configuration after omitting the organic molecules. Consequently, we are able to use a DJ-type parent structure with TPSG *P*4/*mmm* for every child structure. For frameworks with multiple symmetry-inequivalent layers, we would analyze each layer separately, though this was not an issue in any of the structures analyzed herein.

### 3.2. Child structure selection for SMA

The nine child structures selected for the present SMA analysis (listed in **Table 2**) have $PbI_4$ frameworks and organic alkylammonium chains of varying lengths, and were all chosen from a systematic study by Hansen *et al.*,[16] where the reduced exciton mass ($\mu$) of each HOIP compound was calculated by DFT using the generalized gradient approximation (GGA) proposed by Perdew-Burke-Ernzerhof (PBE).[53] The DFT methods employed in the present work are almost identical to those of Hansen *et al.*, as explained above in the Methods section. As the chemical composition of the inorganic framework also affects the band structure, we limit our analysis to lead-iodide based HOIP structures.[55-57] Hansen *et al.*[16] considered HOIP compounds based on $PbCl_4$, $PbBr_4$, and $SnI_4$ frameworks, though there were too few of each framework composition to explore a structural trend. Other studies of HOIP compounds providing DFT values of $\mu$ were considered but not included here because there were too few instances of any given framework composition to explore a structural trend or because the calculations involved different DFT methods.[15,17,27,58,59]

**Table 2.** Single-layer $PbI_4$-based HOIP compounds selected for the present work

| Structure Name | Cation Chemistry | Chemical Formula and Ref. | Structure Identifier |
|---|---|---|---|
| $(BA)_2$@293K | Butylammonium | $CH_3(CH_2)_3NH_3$[60] | CSD 2018895 |
| $(BA)_2$@100K | Butylammonium | $CH_3(CH_2)_3NH_3$[60] | CSD 2018893 |
| C7 | Heptylammonium | $CH_3(CH_2)_6NH_3$[61] | CCDC 805427 |
| C10@172K | Decylammonium | $CH_3(CH_2)_9NH_3$[61] | CCDC 805435 |
| C10@268K | Decylammonium | $CH_3(CH_2)_9NH_3$[61] | CCDC 805436 |
| C14 | Tetradecylammonium | $CH_3(CH_2)_{13}NH_3$[62] | CCDC 692953 |
| 1F-PEPI | 1-F-phenethylammonium | $C_8H_6F_5N$[63] | CCDC 1893385 |
| $(PEA)_2$ | Phenethylammonium | $C_6H_5(CH_2)_2NH_3$[60] | CSD 2018897 |
| EAOH | Ethanolammonium | $HO(CH_2)_2NH_3$[64] | CSD 237189 |

### 3.3 Preparation of child and parent structures for SMA

A unique DJ-type parent structure was prepared for each individual child structure, with an example shown in **Table 3**. These parent structures differ from one child to the next in only three respects. (1) The in-plane unit cell parameter (lattice constant) of the parent lattice is set equal to twice the average in-plane metal-halide bond length of the child. (2) The out-of-plane unit cell parameter is set equal to that of the child – the actual inter-layer spacing has no bearing whatsoever on the symmetry-mode analysis of the inorganic framework. (3) The apical metal-halide separation is set equal to the average apical metal-halide bond length of the child. Furthermore, the out-of-plane axis of the tetragonal parent cell is always set to be the *c*-axis.

**Table 3.** Parent and child framework structures of EAOH $PbI_4$ from Mercier *et al.*[64]

| | a [Å] | b [Å] | c [Å] | $\alpha$ [°] | $\beta$ [°] | $\gamma$ [°] | Atom | x | y | z |
|---|---|---|---|---|---|---|---|---|---|---|
| Parent | 6.461 | 6.461 | 10.214 | 90 | 90 | 90 | Pb1 | 0 | 0 | 0 |
| | | | | | | | I1 | 0 | 0.5 | 0 |
| | | | | | | | I2 | 0 | 0 | 0.3110 |
| Child | 8.935 | 9.056 | 10.214 | 90 | 100.26 | 90 | Pb1 | 0 | 0 | 0.5 |
| | | | | | | | I1 | 0.5694 | 0.4567 | 0.8139 |
| | | | | | | | I2 | 0.7042 | 0.2087 | 0.4799 |

Six of the selected child structures had space groups that were naturally subgroups of the DJ TPSG. The other three structures, $(BA)_2$@293K, C10@268K, and C14, each had two framework layers per unit cell, with interlayer stackings that were close to the RP stacking type and had space groups that were naturally subgroups of the RP TPSG but not the DJ TPSG. These three structures required pre-processing in preparation for one-layer SMA: the manual reduction of the symmetry to space group $P1$, the removal of the "middle" framework layer (i.e. the framework layer closest to (0,0,1/2)), and the detection of the resulting space-group symmetry using FINDSYM.[65] In each of these cases, the omitted middle layer was symmetry-equivalent to the layer that was kept, so that no relevant single-layer distortion data was lost.

### 3.4 Mode descriptions and correlations

In performing SMA on the nine $PbI_4$ alkylammonium compounds studied in the present work, we find three distinct octahedral tilting modes ($M_3^+$, $M_5^+$, and $\Gamma_5^+$), along with two additional octahedral symmetry modes ($\Gamma_1^+$ and $M_2^+$). Since we employ a DJ parent for our SMA, these are modes belonging to irreps of space group $P4/mmm$.[44] All displacive symmetry modes are defined as patterns of linear atomic displacements (i.e. straight-line motions), so that octahedral-tilt modes are linear displacements approximating an octahedral tilt. As mentioned earlier, McNulty and Lightfoot[44] and Liu *et al.*[45] conducted systematic studies of displacive symmetry modes responsible for cooperative octahedral tilting in HOIP frameworks, including the determination of which octahedral tilting modes occur most commonly across known 2D HOIP structures. The structurally-active octahedral-tilt modes we observe are among the most common symmetry modes previously reported. [44,45] It should be noted that Liu *et al.*[45] used an RP parent, which accounts for interlayer correlations, so that their $X$-point irreps of I4/$mmm$ span the same one-layer distortion space as the present $M$-point irreps of $P4/mmm$. As previously noted, we expect interlayer correlations to have negligible impact on the exciton reduced mass, so a single-layer analysis using the DJ parent is sufficient for the present work. Descriptions of each symmetry mode observed in this work are found in **Table 4.** Mode illustrations are also in **Figure 2-4**.

Any two child structures might, in principle, exhibit distortions belonging to very different irreps of the parent, and thus belong to entirely distinct space groups. However, as was recently pointed out by Liu *et al.*,[45] the most common space group for $n = 1$ HOIP structures is $P2_1/c$ (#14), which is the space group of six out of the nine structures considered in this work. The other space groups represented in this analysis are $P\bar{1}$ (#2) and $Pbca$ (#61), which also appear somewhat frequently in the study by Liu *et al.*[45] Each HOIP structure belonging to one of these three space groups, which represents a large fraction of known HOIP structures, is likely to have the structurally active modes of Table 4, among other potential modes. Thus while the structures analyzed in this work represent a small fraction of known HOIPs, we expect the results we obtain to extend to larger families of HOIP structures. One can easily determine which displacive symmetry modes contribute to a given child structure using the free and publicly available ISODISTORT software package (iso.byu.edu).[31,32]

**Table 4.** Description of symmetry modes based on a DJ (*P*4/*mmm*) parent

| Mode symbol | Symmetry-mode description |
| --- | --- |
| $\Gamma_1^+(a)$ | Apical-halide displacement along the parent/child <001> axis, which elongates octahedra along that axis. Equivalent to the application of irrep $\Gamma_1^+(a)$ of *I*4/*mmm* to a single layer. |
| $\Gamma_5^+(a,a)$ | FR tilt pattern with rotational moments along a parent <110> (child <100>) axis that displaces apical halides along the perpendicular in-plane axis. Equivalent to the application of irrep $\Gamma_5^+(a,a)$ of *I*4/*mmm* to a single layer. |
| $M_2^+(a)$ | Checker-board pattern of in-plane Jahn-Teller distortions, which displace equatorial halide atoms along the parent <100> (child <110>) axes. Equivalent to the application of global/site irrep $X_2^+/B_u^2$ $(a,a)$ of *I*4/*mmm* to a single layer. |
| $M_3^+(a)$ | Checkerboard AFR pattern of in-plane cooperative octahedral tilts around the parent/child <001> axis, which displaces equatorial halide atoms along the parent <100> (child <110>) axis. Equivalent to the application of global/site irrep $X_2^+/B_u^3$ $(a,a)$ of *I*4/*mmm* to a single layer. |
| $M_5^+(a,0)$ | A pair of modes that jointly induce a cooperative AFR octahedral tilt pattern of rotational moments along a parent <110> (child <100>) axis. One mode displaces apical halides along the perpendicular in-plane axis, while the other mode displaces equatorial halide atoms along the parent/child <001> axis, and has a corrugated overall appearance. The two modes can be added together in either cooperative or anti-cooperative fashion as shown in Figure 4(c) and (d), though cooperative action tends to be energetically favorable. Equivalent to applying the irrep superposition $X_3^+(a,0)$ $X_4^+$ $(0,b)$ of *I*4/*mmm* to a single layer. |

**Figure 2.** *Graphical representation of the (a) $\Gamma_1^+$ and (b) $\Gamma_5^+$ symmetry-modes acting on a previously undistorted parent structure (faded octahedra) as viewed from within the plane. In (a), the $\Gamma_1^+$ mode displaces apical halides parallel to their parent bonds.*

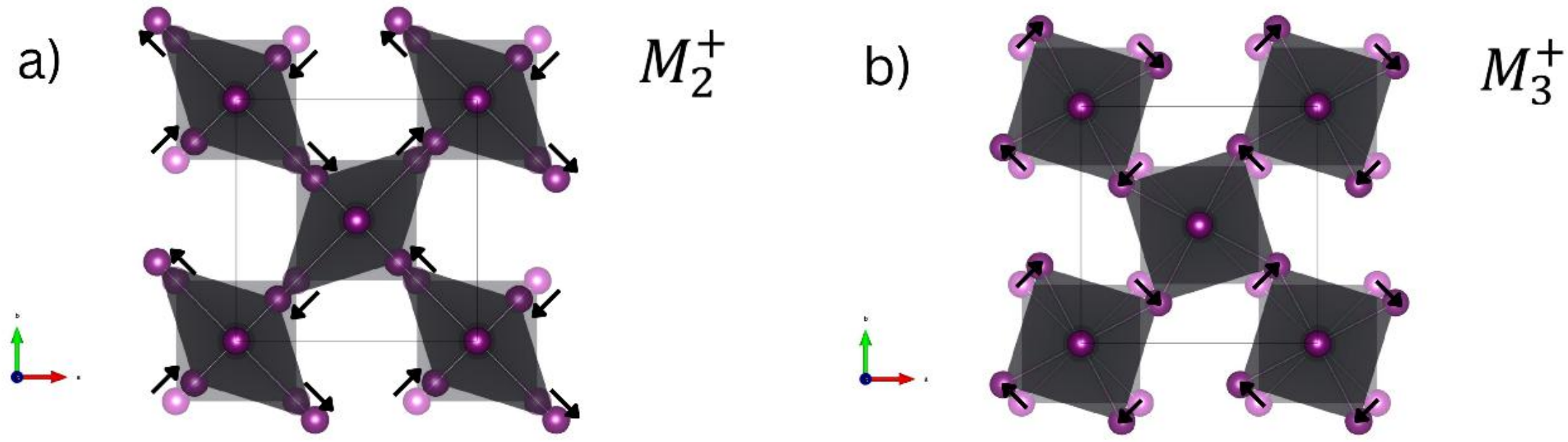

**Figure 3.** *Graphical representation of the (a) $M_2^+$ and (b) $M_3^+$ irreps acting on a previously undistorted parent framework (faded octahedra) as viewed from above the plane.*

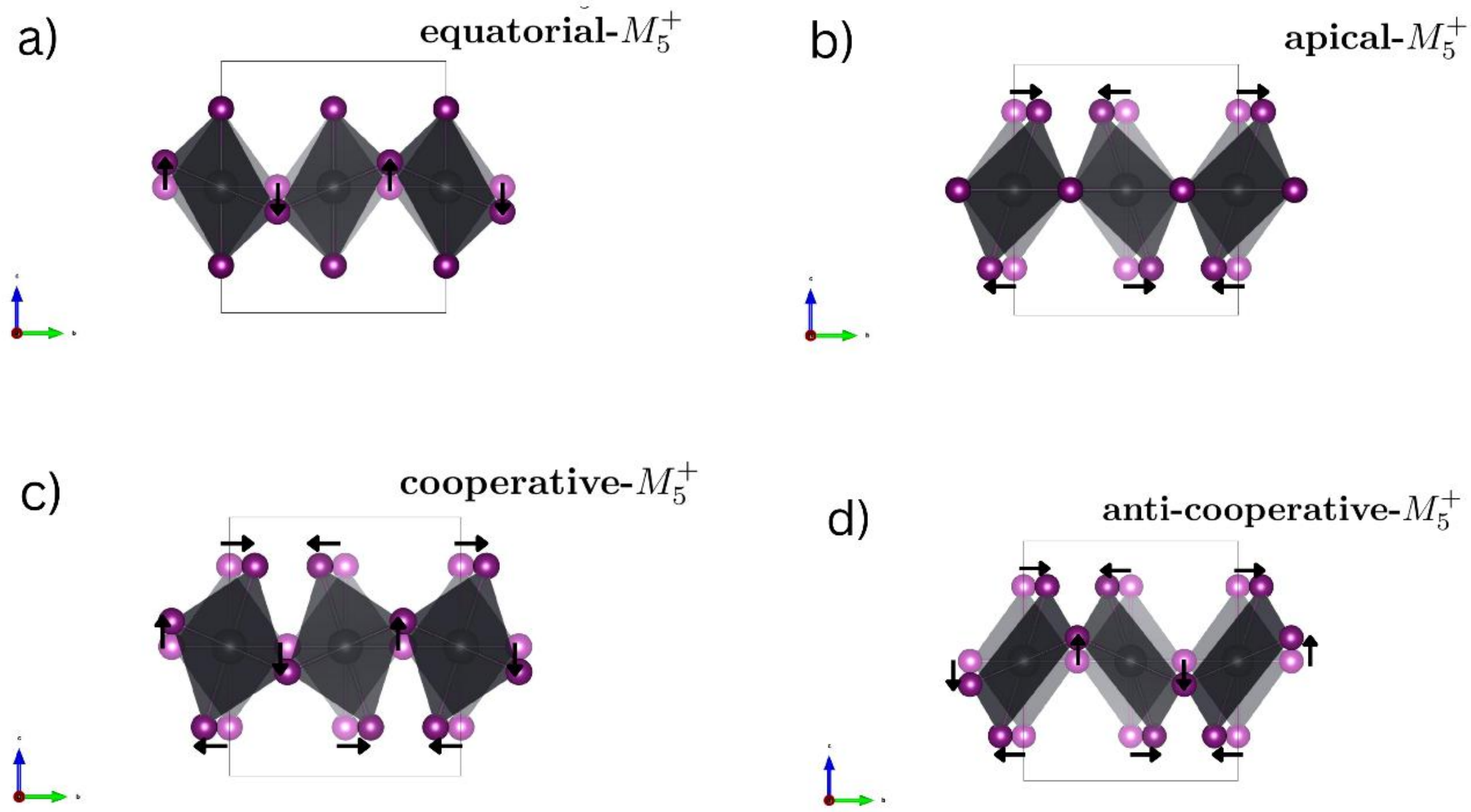

**Figure 4.** *Graphical representation of the $M_5^+$ irrep acting on a previously undistorted parent framework (faded octahedra) as viewed from within the plane. The $M_5^+$ irrep contributes two symmetry modes: (a) one affecting equatorial-halide atomic positions, and (b) the other affecting apical-halide atomic positions. These two symmetry modes can be added together in either a (c) cooperative or (d) anti-cooperative fashion.*

**Table 5** presents the displacive symmetry-mode amplitudes for the nine structures studied in the present work. The symmetry modes are not only correlated to the exciton reduced mass ($\mu$) and the child unit cell parameters, but also to one another, as determined by a correlation analysis, with Spearman- $\rho$ coefficients illustrated in **Figure 5**. Spearman- $\rho$ was used over Pearson- $\rho$ to limit the impact of outliers. For example, if a structure has the $\Gamma_1^+$ displacive mode activated in the positive direction, then it likely has the apical $M_5^+$ displacive mode activated in the positive direction, the $M_3^+$ displacive mode activated in the negative direction, and it also likely has an increased $\mu$ and a decreased "*a*" unit cell parameter, relative to the other materials. We observe a positive correlation between $\mu$ and $\Gamma_1^+$, $M_2^+$, and each of the $M_5^+$ symmetry modes, and observe a negative correlation between $\mu$ and the $\Gamma_5^+$ symmetry mode. The $M_3^+$ symmetry mode has no observable correlation with $\mu$.

The high degree of correlation between the symmetry modes makes it difficult to determine which modes impact $\mu$, and which are merely correlated to impactful symmetry modes.

**Table 5.** *The symmetry mode amplitudes, μ values, and unit cell parameters, for each structure. For compactness, the following mode labels were used: apical-$M_5^+$ → $M_5^+$ (A), and equatorial-$M_5^+$ → $M_5^+$ (E). The $\Gamma_5^+$ and $M_5^+$ modes can be considered as rotational distortions about orthogonal in-plane axes. The last three rows of the table refer to these axes and to the out-of-plane axis.*

| | 1F-PEPI | BA2@100 | BA2@298 | C10@172 | C10@268 | C14 | C7 | $EAOH_2$ | $PEA_2$ |
|---|---|---|---|---|---|---|---|---|---|
| $\Gamma_1^+$ | 0.046 | 0.112 | 0.022 | 0.089 | 0.084 | 0.085 | 0.096 | 0.037 | 0.082 |
| $\Gamma_5^+$ | 0.873 | 0.379 | 0.180 | 0.210 | 0.250 | 0.244 | 0.256 | 0.896 | 0.959 |
| $M_2^+$ | 0.017 | 0.055 | 0.016 | 0.109 | 0.031 | 0.028 | 0.104 | 0.041 | 0.003 |
| $M_3^+$ | 1.077 | 0.915 | 0.982 | 0.893 | 0.947 | 0.927 | 0.918 | 0.795 | 1.065 |
| $M_5^+(A)$ | 0.220 | 0.998 | 0.424 | 0.977 | 0.850 | 0.858 | 1.046 | 0.559 | 0.242 |
| $M_5^+(E)$ | 0.004 | 0.860 | 0.163 | 0.857 | 0.745 | 0.746 | 0.873 | 0.291 | 0.043 |
| μ | 0.115 | 0.139 | 0.111 | 0.146 | 0.134 | 0.135 | 0.147 | 0.122 | 0.108 |
| $M_5^+$ axis | 8.633 (b) | 8.422 (b) | 8.693 (c) | 8.432 (a) | 8.487 (c) | 8.516 (c) | 8.588 (a) | 8.934 (a) | 8.736 (a) |
| $\Gamma_5^+$ axis | 8.799 (c) | 8.995 (a) | 8.876 (b) | 8.949 (b) | 8.831 (b) | 8.847 (b) | 8.941 (b) | 9.056 (b) | 8.737 (b) |
| Out-axis | 16.72 (a) | 26.08 (c) | 27.61 (a) | 21.33 (c) | 43.94 (a) | 54.16 (a) | 17.24 (c) | 10.21 (c) | 16.65 (c) |

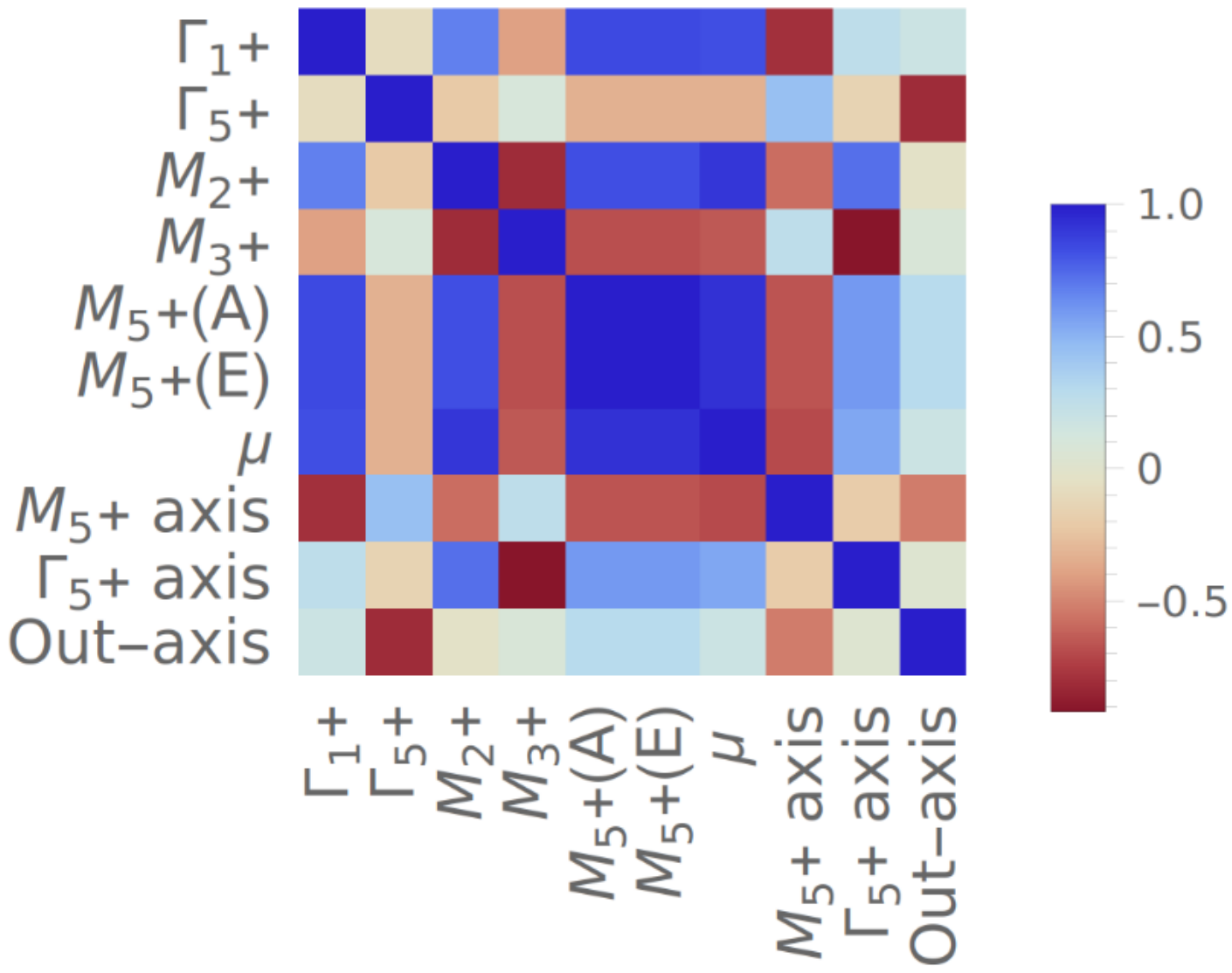

***Figure 5.*** *Correlation heat map for structural and exciton-reduced-mass parameters from Table 5, where blue indicates positive correlation values and red indicates negative correlation values. The $M_5^+$ modes have the same naming convention as in Table 5. The color of each square in the heat map indicates the Spearman-ρ coefficient for the corresponding rows of Table 5, the numerical value of which can be found in the HTML file located in the SI.*

### 3.5 DFT-calculated reduced-mass trends in symmetry-mode parameter space

The positive trend in exciton reduced mass reported by Hansen *et al*. used an octahedral-tilting angle, which was referred to as $|\beta\text{-}\beta'|$, and is qualitatively most similar to the cooperative $M_5^+$ mode, though this bond-angle measure is affected by the anti-cooperative $M_5^+$, $\Gamma_5^+$, and $M_2^+$ modes, among others. From Figure 5 we see that while the cooperative $M_5^+$ mode clearly correlates with $\mu$, several other modes also have a clear correlation with $\mu$. Additionally, the cooperative $M_5^+$ mode is comprised of two distinct symmetry modes (Table 4), each of which could, in principle, impact the band structure in distinct ways. The correlation between symmetry modes makes distinguishing causation from correlation difficult, if not impossible for a small set of structures.

To better understand which symmetry-mode framework-distortions cause the observed changes in $\mu$, we used density functional theory (DFT) to systematically explore the band structure in the region of distortion space surrounding the thermodynamic equilibrium structure of HOIP compound $(EAOH)_2PbI_4$. The equilibrium $(EAOH)_2PbI_4$ structure we employ here was taken from Hansen *et al.*[16], and was obtained by DFT relaxation of a structure determined from single-crystal X-ray diffraction. This HOIP compound was chosen due to a low number of atoms per unit cell, which reduces computational time and complexity, and because the symmetry mode amplitudes of this structure fall roughly in the middle of the ranges typical of its structure family. We note that the $(EAOH)_2PbI_4$ structure has an exceptionally low inter-layer spacing, though this does not affect $\mu$.

For each symmetry mode listed in Table 4, DFT calculations were performed on a sequence of edited $(EAOH)_2PbI_4$ structures, where the mode amplitude was manually fixed at a different value near the equilibrium value. We refer to these structures whose symmetry-mode amplitudes have been manually changed as *DFT-input structures*. Additionally, the nine structures from which Table 1 and Table 2 were constructed, which remain at their thermodynamic equilibria, are referred to as *experimental structures*. Following a procedure similar to that of Dyksik *et al.*,[15] we did not relax the framework or the organic molecules of the DFT-input structures prior to performing the DFT band-structure calculations. We note that each sequence of DFT-input structures contains the *experimental structure* of $(EAOH)_2PbI_4$, which we designate as the *reference structure*.

In an octahedral framework, the approximation of an octahedral tilt by a pattern of linear halide-atom displacements increases the constituent metal-halide bond lengths of the affected octahedra unless accompanied by a compensating lattice strain in the plane perpendicular to the rotation axis. Generally, changing the metal-halide bond lengths may impact the band structure in non-negligible ways.[8] Therefore, the octahedral-tilt modes of $\Gamma_5^+$, $M_3^+$, and $M_5^+$ need strain compensation to preserve the framework metal-halide bond lengths. Because real cooperative polyhedral tilts in framework structures are normally compensated by lattice strains, we also employ lattice-strain compensation in our DFT-input structure sequences to keep the metal-halide bond lengths as close as possible to their original values. In order to judge the impact of strain compensation and bond-length changes on the band structure, we separately analyze DFT-input structure sequences that have no strain compensation. Thus for a given sequence of DFT-input structures, we have generally included two variants per DFT-input structure for all but the reference structure.

For octahedral-tilt modes affecting only the equatorial-halide atom positions ($M_3^+$ and $M_5^+$), compensating strains were applied in directions perpendicular to the axes of rotation. The $M_3^+$ tilt mode rotates equatorial halide atoms around the parent/child <001> axis in an anti-ferrorotational (AFR) pattern and requires a compensating strain along both in-plane axes. The equatorial-atom $M_5^+$ tilt mode rotates equatorial halide atoms around a parent <110> (child <100>) axis in an AFR pattern and requires negative strain compensation along the perpendicular child <100> axis. It was unnecessary to consider strain components that affect the interlayer spacing.

Because the $\Gamma_5^+$ and apical-atom $M_5^+$ tilt modes (as described in Table 4 and illustrated Figure 2b and 4b, respectively) change the in-plane coordinates of apical atoms and thereby increase the corresponding bond lengths, we subsequently renormalize these bond lengths without strain compensation to match those of the reference structure. This renormalization results in an apical displacement that follows a circular arc as opposed to the linear displacements of Figure 2b and 4b. This approach is only possible for apical-atom-only distortions, as equatorial-atom distortions are inherently coupled to the in-plane unit cell parameters.

Some symmetry modes inherently change the metal-halide bond lengths in a way that cannot be corrected by strain compensation, so that including strain-compensated DFT-input structure sequences was not relevant. This is the case for the $\Gamma_1^+$ and $M_2^+$ symmetry modes. The $\Gamma_1^+$ mode is an apical-halide bond-stretching motion, which cannot be compensated for by lattice strain without changing the interlayer spacing. The $M_2^+$ mode is an in-plane Jahn-Teller distortion, which changes metal-halide bond lengths in a way that is impossible to compensate. We observe that these two modes have little to no impact on the DFT-calculated effective mass, as shown in **Figure 6** below.

The figures below plot the $\mu$ values obtained from the present DFT band-structure calculations as a function of symmetry mode amplitudes, with the $\mu$ obtained by the DFT calculations of Hansen *et al*.[16] included for comparison. The *experimental structures* are shown in green (labeled "equilibrium") while *DFT-input structure* sequences are shown in dark blue (strain compensated) and light blue (non strain compensated, labeled "linear"), and the *reference structure* is in red. Trends amongst the DFT-input structure sequences are referred to as *local trends* in the sense that they explore the space around the equilibrium point of a single compound, $(EAOH)_2PbI_4$ (red dot), while trends amongst the experimental structures are referred to as *global trends* in the sense that a broad range of HOIP compounds are represented. Graphs are grouped together as follows, and copies of **Figure 6**, **7**, **8**, and **9** with all structures labeled are available in the SI (**Figure S1, S2, S3, and S4)**:

- Figure 6 – Halide-atom displacements along metal-halide bond axes ($\Gamma_1^+$, $M_2^+$).
- Figure 7 – Apical-halide displacements perpendicular to metal-halide bond axes ($\Gamma_5^+$, $M_5^+$).
- Figure 8 – Equatorial-halide displacements perpendicular to metal-halide bond axes ($M_5^+$, $M_3^+$).
- Figure 9 – Cooperative and anti-cooperative octahedral tilts around in-plane axes ($M_5^+$).

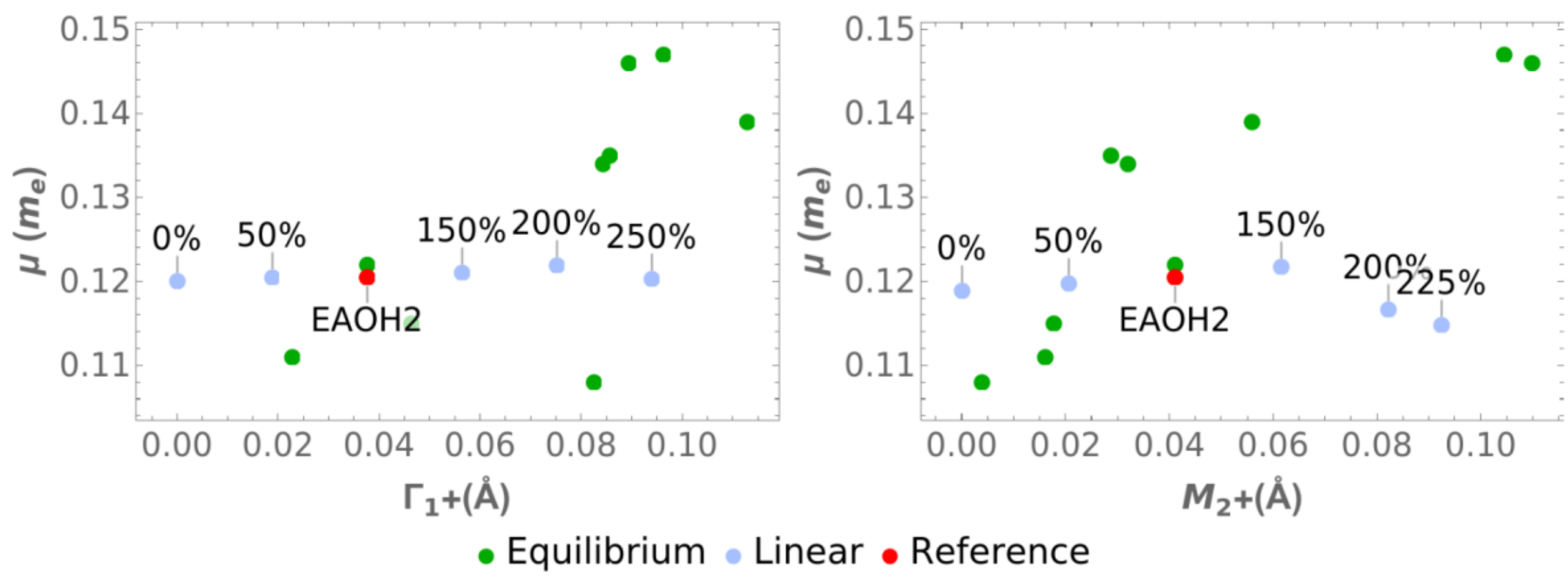

**Figure 6.** *Plot of reduced exciton mass (μ) versus symmetry-mode amplitude (measured in Angstroms) for the $\Gamma_1^+$ (left) and Jahn-Teller (right) $M_2^+$ modes. Both $\Gamma_1^+$ and $M_2^+$ displacements are parallel to parent bonds. Figures 6-9 use the same color convention as follows. Green circles indicate mode amplitudes from the experimental structures, with DFT-*

*simulated μ values obtained from Hansen* et al.[16] *The mode amplitude of the red circle is taken from the* $(EAOH)_2PbI_4$ *structure from Mercier* et al.,[64] *with the μ value coming from current DFT simulations. Light-blue circles have manually-varied mode amplitudes relative to the* $(EAOH)_2PbI_4$ *reference structure (indicated as a percentage of the equilibrium value, not strain-compensated) and μ values are obtained from current DFT simulations. For subsequent figures, strain-compensated structures are indicated by dark-blue circles.*

The $\Gamma_1^+$ and $M_2^+$ distortions cause halide-displacements along metal-halide bond axes. Their global trends (in green) of reduced exciton mass *vs.* mode amplitude (shown in Figure 6) reveal a positive correlation, as expected from Figure 5. However, the local trends (in blue) for the same two modes show very little effect on the band structure, though there may be a slight negative correlation between $M_2^+$ and $\mu$ at high amplitude. This result strongly suggests that these two modes are not the cause of the observed increase in $\mu$

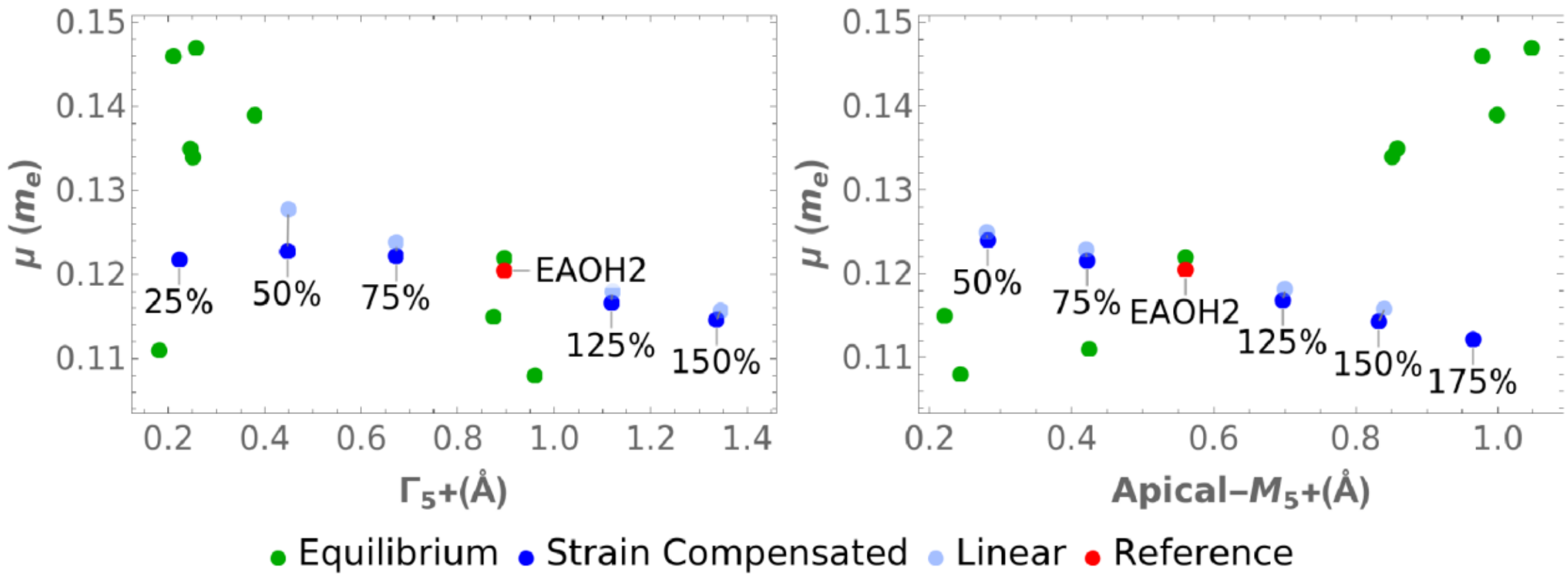

**Figure 7.** *Plot of the reduced exciton mass (μ) against the symmetry mode amplitudes* $\Gamma_5^+$ *(left) and apical* $M_5^+$ *(right), both of which displace apical halide atoms perpendicular to their parent metal-halide bond. Point labels and circle colors are interpreted as in Figure 6. Strain compensation in the context of apical-halide distortions is the renormalization of metal-to-apical-halide bond lengths.*

The $\Gamma_5^+$ and apical $M_5^+$ modes in Figure 7 both yield in-plane displacements of the apical halides perpendicular to their metal-halide bond axes. Globally (in green), we see a strongly-negative correlation between $\Gamma_5^+$ and $\mu$, and a strongly-positive correlation between apical $M_5^+$ and $\mu$. Locally (in blue), we see a clearly-negative correlation between $\Gamma_5^+$ and $\mu$, and a weakly-negative correlation between apical $M_5^+$ and $\mu$ and, which implies that these modes are also not responsible for the correlation reported by Hansen *et al*.[16] Surprisingly, these bond-perpendicular apical-halide displacements decrease $\mu$ locally, i.e. within the distortion space of $(EAOH)_2PbI_4$. To the best of the authors' knowledge, these negative trends have not been observed previously, and could play a crucial role in precision materials-engineering.

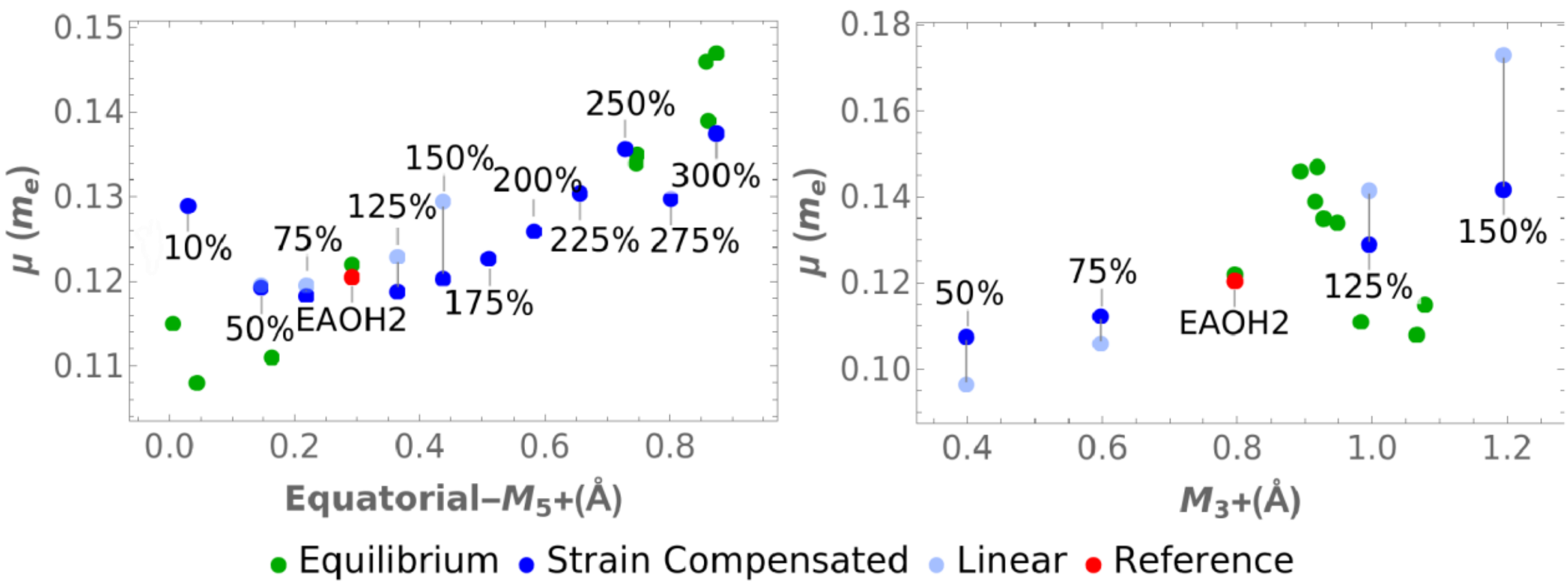

**Figure 8.** *Plot of the reduced exciton mass (μ) against the symmetry mode amplitudes equatorial $M_5^+$ (left) and $M_3^+$ (right), both of which displace equatorial halide atoms perpendicular to their parent metal-halide bond. Point labels and circle colors are interpreted as in Figure 6.*

The equatorial $M_5^+$ and $M_3^+$ modes of Figure 8 displace equatorial halide atoms in directions perpendicular to their metal-halide bond axes. We focus first on equatorial $M_5^+$, where we see a clear positive global correlation with $\mu$ (in green). Locally (in blue), the equatorial $M_5^+$ mode has a clearly positive correlation with $\mu$ (except for the lowest mode amplitudes). It is possible that at low mode amplitudes the contributions of the equatorial $M_5^+$ mode are washed out by contributions from other modes, or that the effects from straining the unit cell become non-negligible. However, through the clear agreement between global and local trends of equatorial $M_5^+$, a key structure-property relationship has been revealed: this mode, namely the bond-perpendicular equatorial-halide displacements of Figure 4a—and not the cooperative $M_5^+$ motion of Figure 4c—is likely the structural distortion most responsible for altering the reduced exciton mass in this family of 2D HOIPs

Globally, the $M_3^+$ mode has an unclear effect on $\mu$, though we observe that $M_3^+$ has a large local impact on the band structure for $(EAOH)_2PbI_4$. Dyksik *et al*. reported positive correlations between $\mu$ and a related set of octahedral tilt angles that are similar to both the equatorial $M_5^+$ and $M_3^+$ modes, specifically finding that increasing distortion parameters like $M_3^+$ and equatorial $M_5^+$ affect the hybridization between iodine *p* states and lead *s* states.[15] We note that the standard deviation of the $M_3^+$ mode amplitudes observed in this work is small compared to the average mode amplitude value. From the local trend observed in this work and the conclusions of Dyksik *et al.,* we expect that a larger set of structures would also reveal a positive correlation between $M_3^+$ and $\mu$.

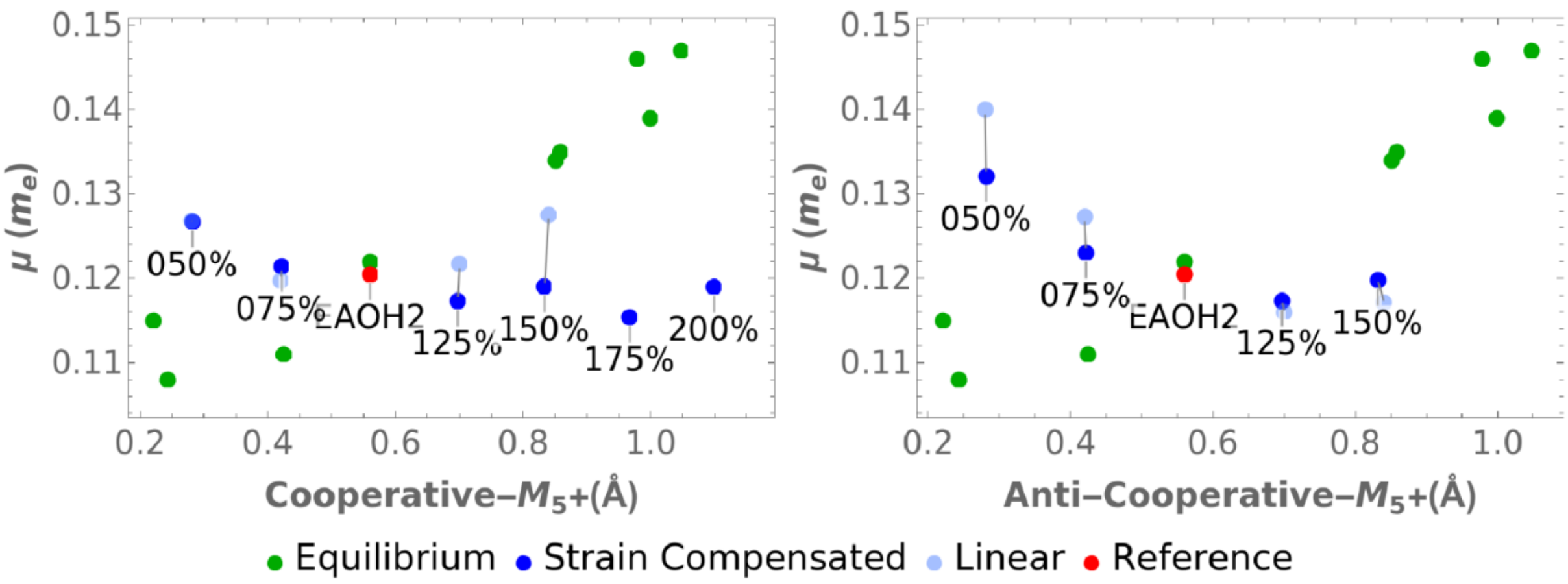

**Figure 9.** *Plot of the reduced exciton mass ($\mu$) against the apical-halide symmetry-mode amplitudes for the cooperative $M_5^+$ (left) and anti-cooperative (right) $M_5^+$ modes. For a given apical-atom mode amplitude, the corresponding equatorial-atom amplitude was chosen to be completely cooperative or completely anti-cooperative. Point labels and circle colors are interpreted as in Figure 6.*

In search of the most intuitive description of the structural features that directly impact μ, we have chosen to employ two distinct bases for the two dimensional $M_5^+$ parameter space: the equatorial and apical $M_5^+$ modes (Figure 4a-b), and the cooperative and anti-cooperative $M_5^+$ modes (Figure 4c-d). While the anti-cooperative mode is energetically unfavorable at higher amplitudes, it provides an orthogonal complement to the cooperative mode in a 2D parameter space, which must be explored fully. To force a distortion to be completely cooperative (unchanged halide-metal-halide bond angles) or anti-cooperative (maximally changed halide-metal-halide bond angles), the apical halide mode (apical $M_5^+$) is first chosen as the independent parameter, and the in-plane $M_5^+$ mode is then made to depend on it in the appropriate way.

The cooperative $M_5^+$ symmetry mode of Figure 9 is the most similar to the pattern of octahedral tilts that Hansen *et al.*[16] reported to correlate with $\mu$. However, the local $\mu$ trends of both the cooperative $M_5^+$ and anti-cooperative $M_5^+$ sequences (in blue) are roughly parabolic – not closely similar to the observed global $\mu$ trend across the entire family of structures (in green) – indicating that neither the cooperative $M_5^+$ mode nor the anti-cooperative $M_5^+$ mode is likely to be the cause of the observed global $\mu$ trend.

The landscape illustrated in **Figure 10**, which combines the strain-compensated structures from each of the $M_5^+$ modes and mode combinations, shows a non-linear dependence on the apical $M_5^+$ and equatorial $M_5^+$ mode amplitudes. The "apical" and "equatorial" axes of the plot are the mode amplitudes from the apical $M_5^+$ and equatorial $M_5^+$ modes respectively. The cooperative $M_5^+$ and anti-cooperative $M_5^+$ axes lie between the apical and equatorial axes, forming an "X" pattern in the *x-y* plane of Figure 10. Combining the apical $M_5^+$ and equatorial $M_5^+$ together in some fixed proportion within the framework does not necessarily combine their $\mu$ values in the same proportion. This fact makes it difficult to anticipate the behavior of $\mu$ in the interpolated regions of the mode space between the structure sequences we employed. However, we can clearly see from this landscape that the equatorial $M_5^+$ mode has the largest impact on $\mu$.

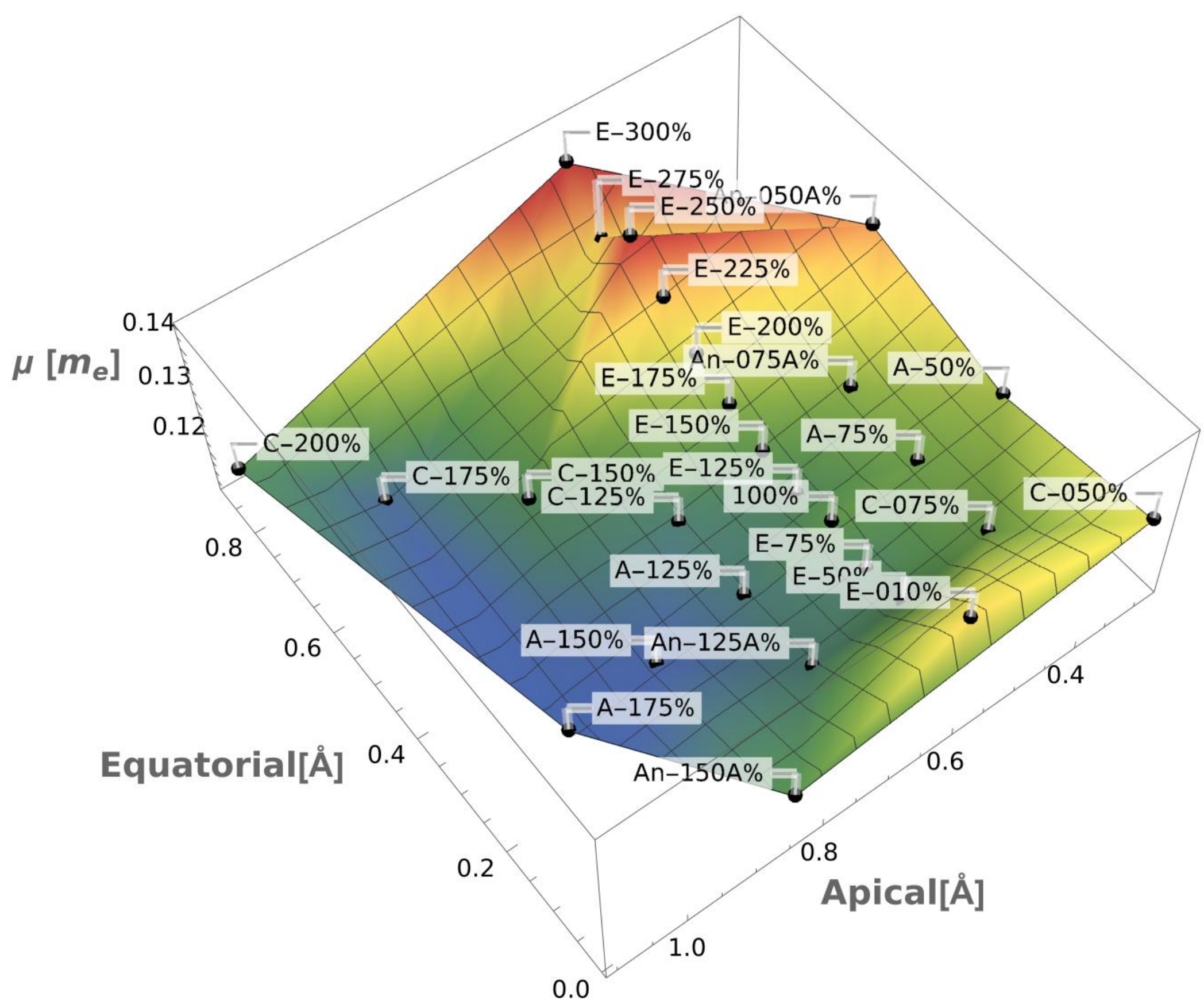

**Figure 10.** *Representation of the reduced exciton mass (μ) landscape of the* $(EAOH)_2PbI_4$ *framework as a function of the equatorial* $M_5^+$ *(E) and apical* $M_5^+$ *(A) symmetry modes from Figure 4a-b). The cooperative* $M_5^+$ *(C) and anti-cooperative* $M_5^+$ *(An) modes from Figure 4c-d) are also plotted along these axes, forming an "X" in the x-y plane. Each black dot is a strain-compensated framework structure corresponding to a dark-blue circle from one of Figures 7(left and right), 8(right) or 9(left).*

## 4. Conclusion

Hybrid organic-inorganic perovskites (HOIPs) provide a rich variety of structure-property relationships, including a recently reported connection between an inorganic octahedral-tilt angle and the reduced exciton mass ($\mu$) near the band gap.[15,16] Specifically, $\mu$ was observed to increase alongside an alternating $PbI_4$ octahedral tilt about an in-plane axis for a series of alkyl-chain and aromatic single-layer HOIP structures. Using group representation theory, we decomposed each distorted child framework into displacive symmetry modes of a common high-symmetry topological parent framework, providing six independent and orthogonal symmetry modes, which each contribute to one or more of the various octahedral-tilt angles purported in the literature to influence the band structure.

Previous studies have used the orthogonal and complete basis of symmetry modes to characterize the possible distortions of perovskite systems[9,36-39,44,45] while others have focused on characterizing the electronic structures of HOIPs.[10,12-14,18-27] Here, we have used DFT calculations to explore band structure as a function of the robust and unambiguous symmetry-mode parameter set, and thereby have been able to distinguish causation from correlation in an important structure-property relation – namely the distortions which influence reduced exciton mass – which has not been accomplished previously.

By independently/separately varying the amplitudes of each of the available displacive symmetry modes around the equilibrium structure of $(EAOH)_2PbI_4$, and by using DFT to calculate the value of $\mu$ for each distorted framework generated, we have identified three structure-property relationships.

1) *Bond-perpendicular equatorial-halide displacements* (equatorial-$M_5^+$ and $M_3^+$) increase $\mu$, with equatorial-$M_5^+$ taking the dominant role among the structures analyzed in this work.
2) *Bond-perpendicular apical-halide displacements* ($\Gamma_5^+$ and apical-$M_5^+$) decrease $\mu$, albeit at a lower rate than achieved by the bond-perpendicular equatorial-halide displacements.
3) *Halide-atom displacements parallel to metal-halide bonds* ($\Gamma_1^+$ and $M_2^+$) don't significantly impact $\mu$.

We expect these three rules to have predictive relevance to other families of single-layered HOIP compounds, even if the vertical biases and trend-slopes differ from family to family.

The DFT exploration of physical- and electronic-property landscapes parameterized by symmetry modes is a promising new avenue for band-structure engineering in the broader family of perovskite materials. In addition to improved control over band gap, exciton binding energy, and reduced exciton mass, DFT/SMA investigations of HOIP-framework compounds may enable us to isolate the root structural causes of advanced electronic properties such as Rashba-Dresselhaus spin-splitting and chirality-induced spin selectivity (CISS).[3,4,20,21,66]

**Acknowledgements**

We acknowledge Ben Fuqua MS for advice and assistance in creating the interactive HTML content in the Supplementary Information.

We acknowledge funding from NSF grant No. 2139185. This work also made use of the Illinois Campus Cluster, a computing resource that is operated by the Illinois Campus Cluster Program in conjunction with the National Center for Supercomputing Applications, which is supported by funds from the University of Illinois Urbana-Champaign. JSC acknowledges partial support from the Agile Electronics Materials and Processes Research Team of the Materials and Manufacturing Directorate, Air Force Research Laboratory.

**Data Availability Statement**

The data that support the findings of this study are either in the supplementary information or are available from the corresponding author upon reasonable request.

**Supporting Information for "Isolating the structural distortions that influence reduced exciton mass via symmetry-mode decomposition and DFT analysis"**

*Isaac R. Burkholder, Cindy Y. Wong, André Schleife, Kameron Hansen, John S. Colton, Branton J. Campbell**

The interactive HTML content in the Supplementary Information (DOI: 10.5281/zenodo.22261753) is based on the framework structures and reduced exciton mass ($\mu$) values from Hansen *et al*.[12] and was created using the YData-profiling python package. MathJax was then manually inserted into the HTML data to enhance the presentation of the data. The interactive HTML files are each split into 4 sections: *Overview*, *Variables*, *Interactions*, and *Correlations*. One can jump to any section using the hyperlinks located in the top-right corner.

The *Overview* section contains general information about the dataset as a whole and is split into three menu items: *Overview*, which shows statistics for the whole dataset, *Alerts*, which points out parameters that are highly correlated or have entirely unique values, and *Reproduction*, which contains the configuration files used to generate the HTML file. Neither type of alert is cause for concern because we are specifically looking for correlations and because each structure is clearly distinct. The highly correlated parameters from the *Alerts* menu item can be easily visualized using the later *Interactions* and *Correlations* sections of the HTML file.

The *Variables* section allows a user to select a variable from a drop-down, which then display specific statistical information about that variable. Clicking the "more details" option at the bottom-right corner of the display opens a panel containing additional statistical information.

The *Interactions* section allows a user to select two variables from drop-down menus and visualize their correlation with a plot. The variable from the left-hand menu varies along the horizontal axis of the plot and the variable from the right-hand menu varies along the vertical axis of the plot.

The *Correlations* section has two menu items: *Heatmap* and *Table*. The *Heatmap* enables the user to succinctly visualize the correlations within each of the variable pairs in a single graphic. The numerical correlation values were calculated using the Spearman- $\rho$ test. The *Table* menu item shows the numerical correlation values from the *Heatmap*.

We also include here copies of Figs. 6-9, with each data-point labeled with the corresponding structure.

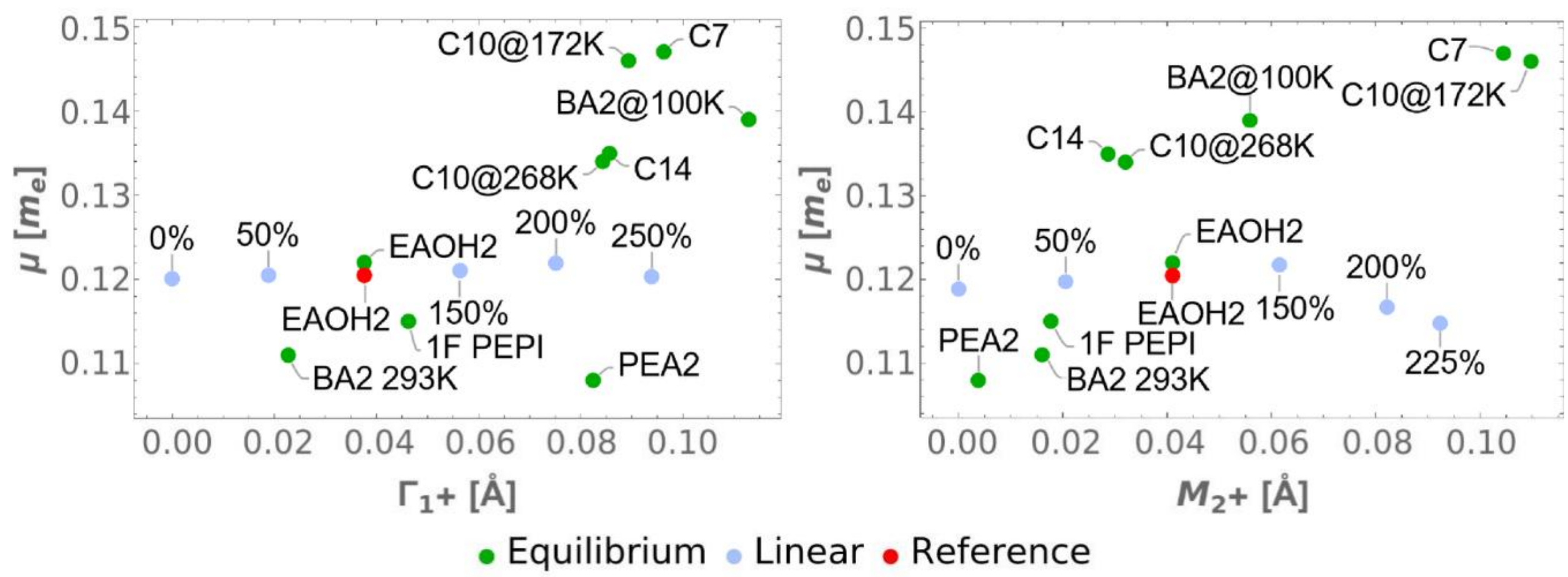

**Figure S1.** *Plot of reduced exciton mass (μ) versus symmetry-mode amplitude (measured in Angstroms) for the $\Gamma_1^+$ (left) and Jahn-Teller (right) $M_2^+$ modes. Both $\Gamma_1^+$ and $M_2^+$ distortions are parallel to parent bonds. Figures S2-S4 use the same color convention, as follows. Green circles indicate mode amplitudes and DFT-simulated μ values extracted from the experimental structures analyzed by Hansen* et al.*[16] The mode amplitude of the red circle is taken from the $(EAOH)_2PbI_4$ structure from Mercier* et al.*,[64] with the μ value coming from current DFT simulations. Dark-blue and light-blue circles have manually-varied mode amplitudes relative to the $(EAOH)_2PbI_4$ reference structure (indicated as a percentage of the equilibrium value) and μ values are obtained from current DFT simulations. Dark-blue and light-blue circles indicate strain-compensated and non-strain-compensated structures, respectively. Strain compensation (dark blue) is not applicable to the modes shown here.*

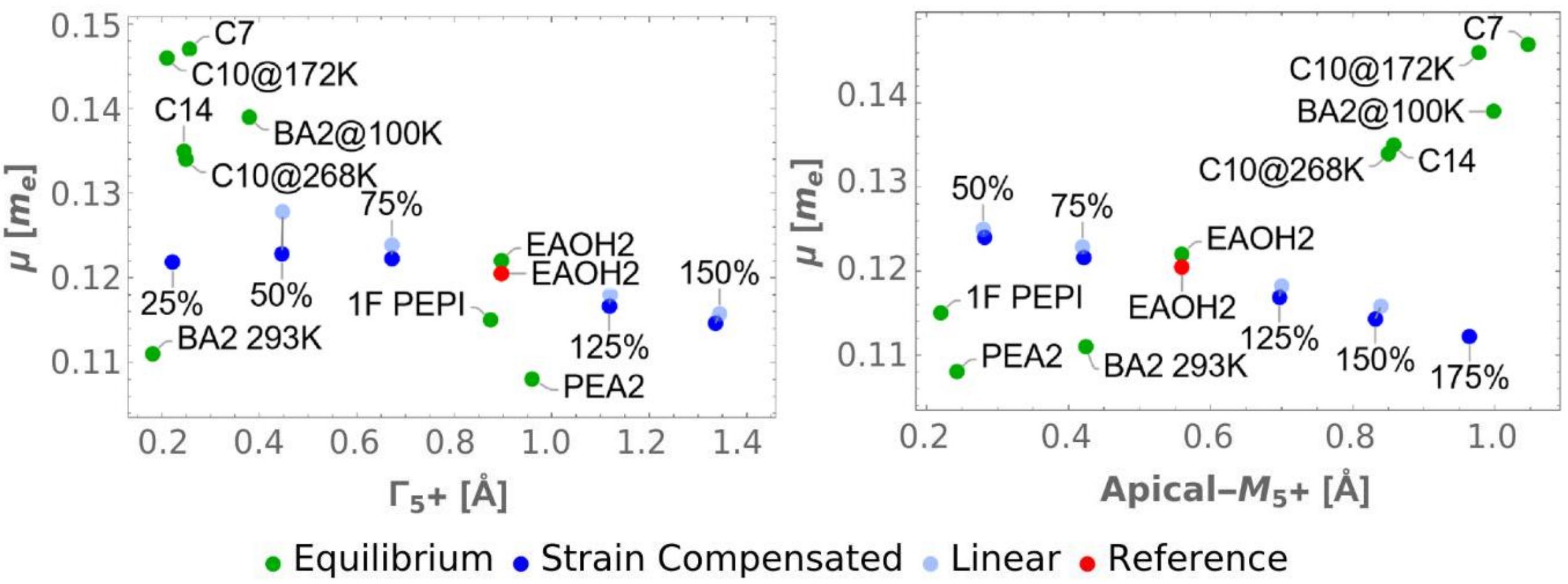

**Figure S2.** Plot of the reduced exciton mass (μ) against the symmetry mode amplitudes $\Gamma_5^+$ (left) and apical $M_5^+$ (right). Both $\Gamma_5^+$ and apical $M_5^+$ displace apical halide atoms perpendicular to their parent metal-halide bond. Point labels and circle colors are interpreted

as in Figure S1.

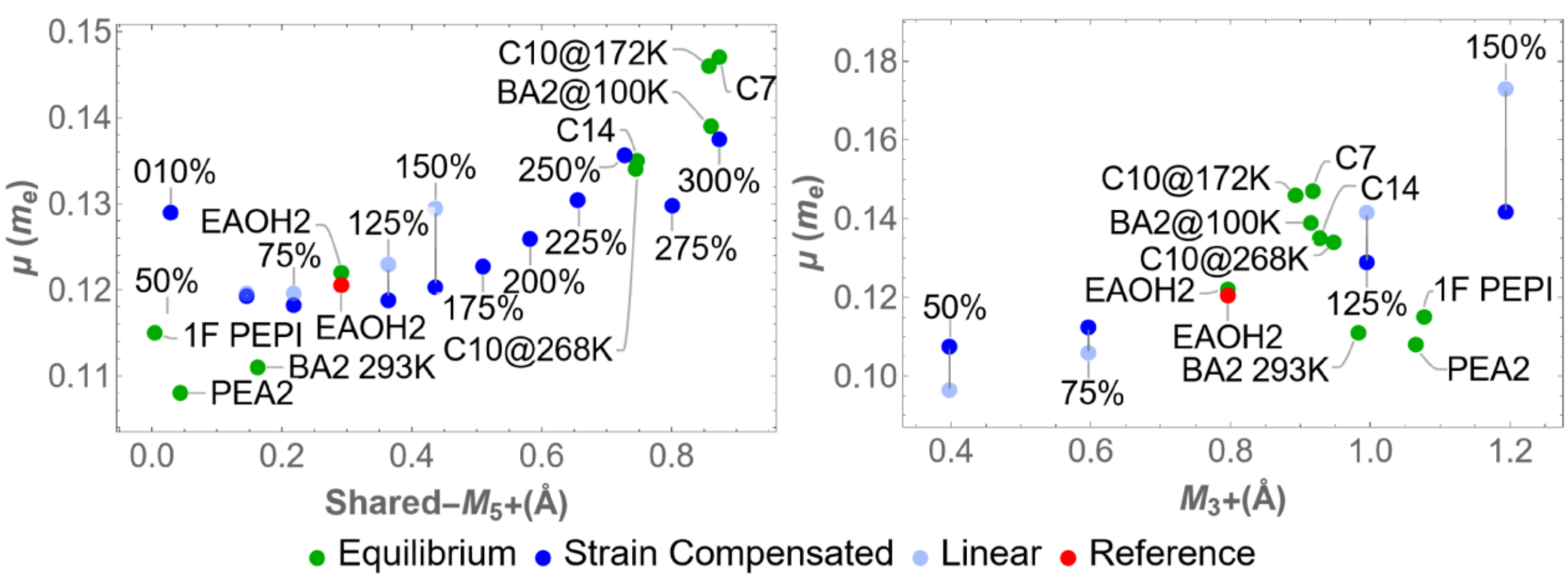

**Figure S3.** Plot of the reduced exciton mass (μ) against the symmetry mode amplitudes equatorial $M_5^+$ (left) and $M_3^+$ (right). Both equatorial $M_5^+$ and $M_3^+$ displace equatorial halide atoms perpendicular to their parent metal-halide bond. Point labels and circle colors are interpreted as in Figure S1.

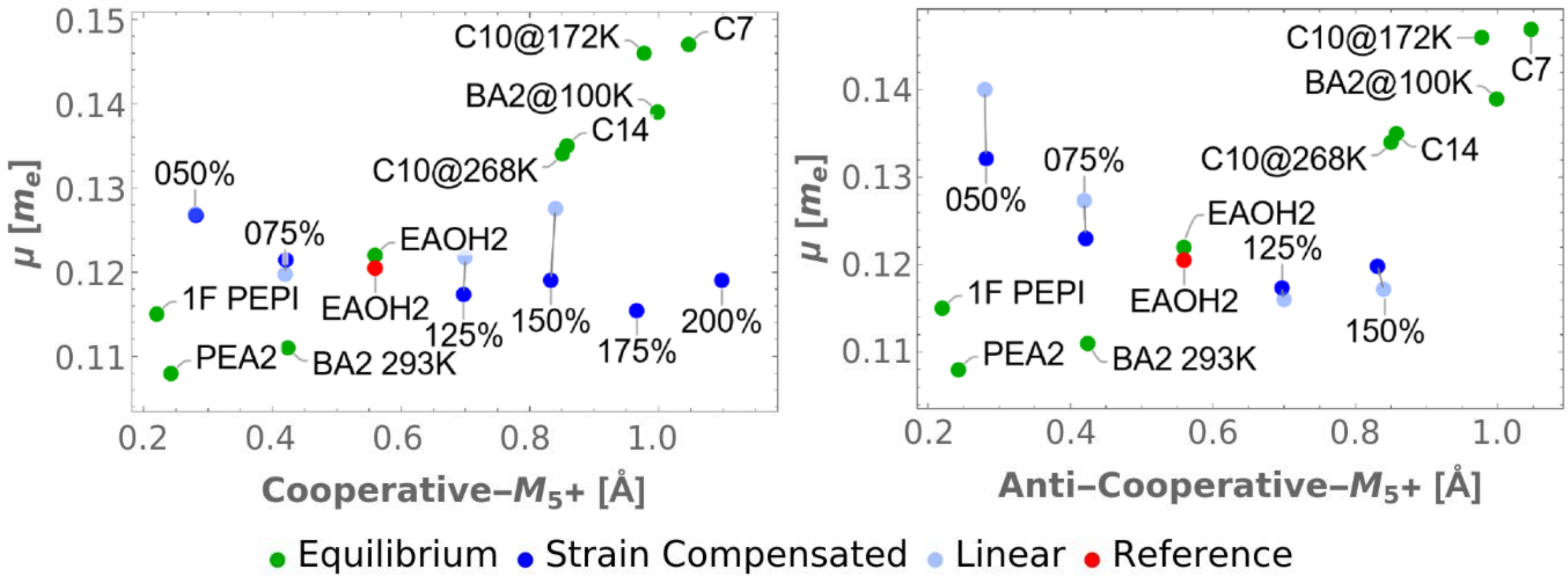

**Figure S4.** Plot of the reduced exciton mass (μ) against the apical-halide symmetry-mode amplitudes for the cooperative $M_5^+$ (left) and anti-cooperative (right) $M_5^+$ modes. For a given apical-atom mode amplitude, the corresponding equatorial-atom amplitude was chosen to be completely cooperative or completely anti-cooperative. Point labels and circle colors are interpreted as in Figure S1.